\documentclass[aps,prd,amsmath,amssymb,superscriptaddress,altaffillsymbol, preprintnumbers,preprint,nofootinbib,a4paper]{revtex4}
\pdfoutput=1
\usepackage{amsthm}
\usepackage{graphicx}
\usepackage{color}
\newcommand{\beq}{\begin{eqnarray}}
\newcommand{\eeq}{\end{eqnarray}}

\newcommand{\bmp}{\noindent\begin{minipage}{16cm}}
\newcommand{\emp}{\end{minipage}\vskip 7mm} 

\usepackage{dcolumn}
\usepackage{bm}
\usepackage{bbm}
\usepackage{subfigure}
\usepackage{pxfonts}
\usepackage{slashed}

\theoremstyle{definition}

\theoremstyle{plain}

\usepackage{epsfig}

\usepackage[ margin=5pt, font=normalsize,labelfont=bf,justification=raggedright]{caption}

\usepackage{hyperref}
\definecolor{rossoCP3}{cmyk}{0,.88,.77,.40}
\definecolor{verdeCP3}{rgb}{0.09765625, 0.57421875, 0.1015625}
\definecolor{bluCP3}{rgb}{0, 0.23, 0.67}
\hypersetup{colorlinks, bookmarksopen, bookmarksnumbered,citecolor=verdeCP3, linkcolor=bluCP3, pdfstartview=FitH, urlcolor=rossoCP3}
\def\lsim{\mathrel{\rlap{\lower4pt\hbox{\hskip1pt$\sim$}}
    \raise1pt\hbox{$<$}}}                
\def\gsim{\mathrel{\rlap{\lower4pt\hbox{\hskip1pt$\sim$}}
    \raise1pt\hbox{$>$}}}                

\newcommand{\bea}{\begin{eqnarray}}
\newcommand{\eea}{\end{eqnarray}}

\newcommand{\ba}{\begin{eqnarray}}
\newcommand{\ea}{\end{eqnarray}}

\newcommand{\be}{\begin{eqnarray}}
\newcommand{\ee}{\end{eqnarray}}

\begin{document}
\title{\Large  \color{rossoCP3} ~~\\ The Up-Shot of Inelastic Down-Scattering at CDMS-Si}
%
\author{Mads T. Frandsen}
\email{frandsen@cp3-origins.net} 
\author{Ian M. Shoemaker}
\email{shoemaker@cp3-origins.net} 
\affiliation{{\color{rossoCP3} CP$^{3}$-Origins} \& Danish Institute for Advanced Study {\color{rossoCP3} DIAS}, University of Southern Denmark, Campusvej 55, DK-5230 Odense M, Denmark}

\begin{abstract}
We study dark matter that inelastically scatters and de-excites in direct detection experiments, as an interpretation of the CDMS-Si events in light of the recent LUX data. The constraints from LUX and XENON10 require the mass-splitting between the DM excited and de-excited states to be $|\delta| \gtrsim 50$ keV. At the same time, the CDMS-Si data itself do not allow for a consistent DM interpretation for mass splittings larger than $|\delta| \sim $200 keV. We find that a low threshold analysis will be needed to rule out this interpretation of the CDMS-Si events.
 In a simple model with a kinetically mixed dark photon, we show that the CDMS-Si rate and the thermal relic abundance can both be accommodated. 
\\[.1cm]
{\footnotesize  \it Preprint: CP3-Origins-2013-053 DNRF90, DIAS-2013-53.}
 \end{abstract}

\maketitle

\section{Introduction}
An unambiguous, non-gravitational signal of Dark Matter (DM) has yet to be detected. However, a number of direct detection experiments have reported low-energy recoil events in excess of known backgrounds. These observations, and their consistency or inconsistency with null experiments, have been widely discussed in the literature. Most recently, the CDMS-II collaboration reported 3 events in Silicon (Si) detectors with an expected background of 0.62 events \cite{Agnese:2013rvf} from an exposure of 140.2 kg-days. Taken in isolation, this data prefers a DM plus background hypothesis over a background only hypothesis with a probability of 99.8$\%$ and the highest likelihood for a DM mass of $\sim 8.5$ GeV and a cross-section of $2\times 10^{-41} \rm{cm}^2$, assuming the standard picture of spin-independent interactions  \cite{Agnese:2013rvf}. ~\footnote{Although we do not explore non-DM explanations of these signals, it has been observed that most of the direct detection anomalies can be accommodated with a model of baryonic neutrinos~\cite{Pospelov:2011ha,Pospelov:2012gm} sourced by ordinary solar neutrinos, even in the aftermath of LUX~\cite{Pospelov:2013rha}.}

It was shown~\cite{Frandsen:2013cna} to be possible to accommodate the null results of the XENON10 \cite{Angle:2011th} and XENON100 \cite{Aprile:2012nq} experiments with a positive CDMS-Si signal, in particular if either DM scatters inelastically to a lower mass state, termed variously ``Exciting'' and``Exothermic'' DM \cite{Finkbeiner:2007kk,Batell:2009vb,Graham:2010ca} in the literature, or if the scattering on protons and neutrons differ, recently termed ``isospin-violating'' DM (IVDM)~\cite{Kurylov:2003ra,Giuliani:2005my,Chang:2010yk,Feng:2011vu}. However, after the appearance of the LUX data \cite{Akerib:2013tjd} 
it was observed~\cite{Gresham:2013mua,DelNobile:2013gba,Cirigliano:2013zta} that LUX and CDMS-Si are now in strong tension even in the case of isospin-violating DM (IVDM) where the ratio of the proton-to-neutron couplings can be maximally ``xenophobic.''  

Exothermic DM (exoDM) models~\cite{ArkaniHamed:2008qn,Batell:2009vb,Graham:2010ca,Essig:2010ye,McCullough:2013jma} have not been considered in the aftermath of LUX. This is the topic of the present paper. 
With a mass-splitting between the DM state and its excited state of $\delta = -50$ keV, we find that strong tension with the LUX/XENON10 results remains. This is significantly improved by going to larger mass splittings, $\delta \sim -200$ keV. Increasing the mass-splitting yet further degrades the ability of exoDM to fit the CDMS-Si data. This is because larger mass-splittings are unable to simultaneously account for the high and low-energy events. With present data LUX does not exclude the best-fit point with a $\delta = -200$ keV mass splitting. 

The DM masses and mass-splittings we find that are needed to accommodate the CDMS-Si data have not been explored in prior work. We therefore construct an illustrative toy model with a kinetically mixed photon that can simultaneously account for the thermal relic abundance, the requisite CDMS-Si signal, and ensure that the higher mass state is well-populated at the present epoch.

\section{LUX Experimental Details}

\begin{figure}[t!]
  \centering
    \includegraphics[width=0.5\textwidth]{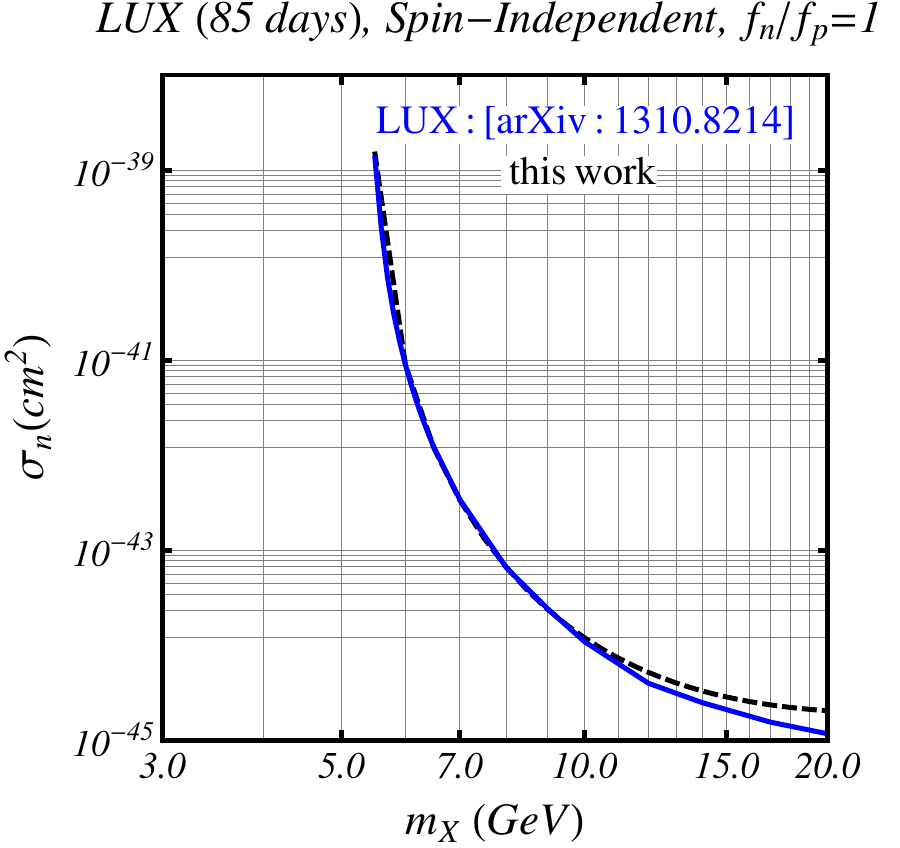}
   \caption{Comparison of our fit to LUX data~\cite{Akerib:2013tjd} (black dashed) against that of the collaboration (blue solid).}
  \label{fig:comp}
\end{figure}

The LUX experimental design is quite similar to XENON100. Both experiments use the combination of scintillation (S1) and ionization signals (S2) to effectively reject background. In the standard analysis done by the xenon-based experiments, it is important to account for the Poissonian nature of the distribution of S1 photoelectrons. As such, we compute the number of signal events following~\cite{Aprile:2011hx} as,
\be
N_{DM} = \int_{S1_{lower}}^{S1_{upper}} dS1 \sum_{n=1}^{\infty} {\rm Gauss}(S1| n, \sqrt{n} \sigma_{PMT}) \int_{0}^{\infty} dE_{R} \varepsilon(E_{R}) ~{\rm Poiss}(n|\nu(E_{R})) \frac{dR}{dE_{R}} \times {\rm Exp},
\ee
where ${\rm Exp}$ is the experimental exposure, $ \varepsilon(S1)$ is the S1 efficiency, $\sigma_{PMT} = 0.37$ PE accounts for the PMT resolution~\cite{Akerib:2012ys}. For the LUX analysis, $S1_{lower}=2$ and $S1_{upper} = 30$.  The expected number of photoelectrons $\nu(E_{R)})$ is
\be \nu(E_{R}) = E_{R} \times \mathcal{L}_{eff}(E_{R}) \times \frac{S_{nr}}{S_{ee}} \times L_{y}
\ee
where $\mathcal{L}_{eff}$ is the energy-dependent scintillation efficiency of liquid Xenon, $L_{y}$ is the light yield, and $S_{nr}, S_{ee}$ are the nuclear and electron recoil quenching factors respectively arising from the applied electric field.  Importantly for LUX's sensitivity to low mass DM, their absolute light yield has been measured to be much larger than XENON100's. We use the energy dependent absolute light yield, $\mathcal{L}_{eff}(E_{R})  \frac{S_{nr}}{S_{ee}}  L_{y}$, taken from slide 28 of the talk~\cite{lux-talk}, with a hard cut-off below 3 keV. Finally for the WIMP detection efficiency in Eq. (1) above we use the efficiency {\it before} threshold cuts (green triangles) taken from Fig. 9 of~\cite{Akerib:2013tjd}. Following the collaboration's treatment of low-mass WIMPs, we compute 90$\%$ CL limits by requiring, $N_{DM} < 2.4$.

For the details of our treatment of XENON10 and CDMS-Si we refer the reader to~\cite{Cirigliano:2013zta}.



\section{Inelastic Interpretations}
Shortly after the CDMS-Si results, it was pointed out that both IVDM and exoDM can account for the CDMS-Si results while remaining consistent with the constraints from XENON10 and XENON100~\cite{Frandsen:2013cna}. However after LUX, interpretations CDMS-Si in terms of elastic DM scattering, with or without isospin-violation seem strongly disfavored~\cite{Gresham:2013mua,DelNobile:2013gba,Cirigliano:2013zta}. For example, a recent study~\cite{Gresham:2013mua} of elastic scattering examined a large number of operators, finding no way of substantially reducing the tension between CDMS-Si and the null searches~\cite{Angle:2011th,Aprile:2012nq,Akerib:2013tjd}. Similarly~\cite{DelNobile:2013gba} concluded in an astrophysics-free manner that elastic, spin-independent scattering is insufficient to significantly reduce the tension. Yet none of these studies have investigated the inelastic interpretation of direct detection data in the aftermath of LUX. Let us now focus on the nature of the inelasticity needed. 

The salient feature of inelastic scattering is that upon scattering with a nucleus, the rest mass of the DM changes, 
\be X_1 +  \mathcal{N} \rightarrow X_{2} +\mathcal{N} 
\ee
The requisite velocity to produce a nuclear recoil of energy $E_{R}$ is
\be v_{\min} = \frac{1}{\sqrt{2E_{R}m_{N}}} \left |  \delta + \frac{m_{N}E_{R}}{\mu} \right|
\ee
where $\delta \equiv m_{2}-m_{1}$ is the mass splitting. Therefore up-scattering ($\delta > 0$) is more favorable on heavy nuclei, whereas down-scattering ($\delta < 0$) is more favorable on light nuclei. Given the lightness of Si with respect to Xe, down-scattering is a way to explain the CDMS-Si data while remaining consistent with the null Xenon searches. 

\begin{figure*}[t!]
  \centering
  \includegraphics[width=0.45\textwidth]{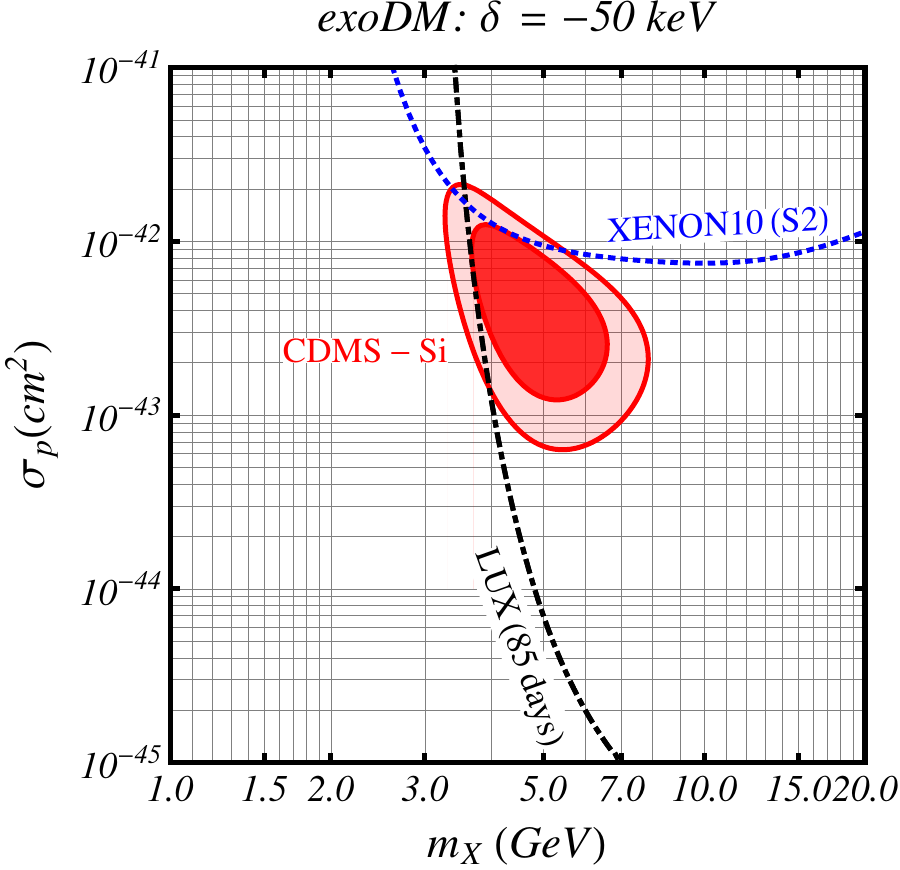}
                 \includegraphics[width=0.45\textwidth]{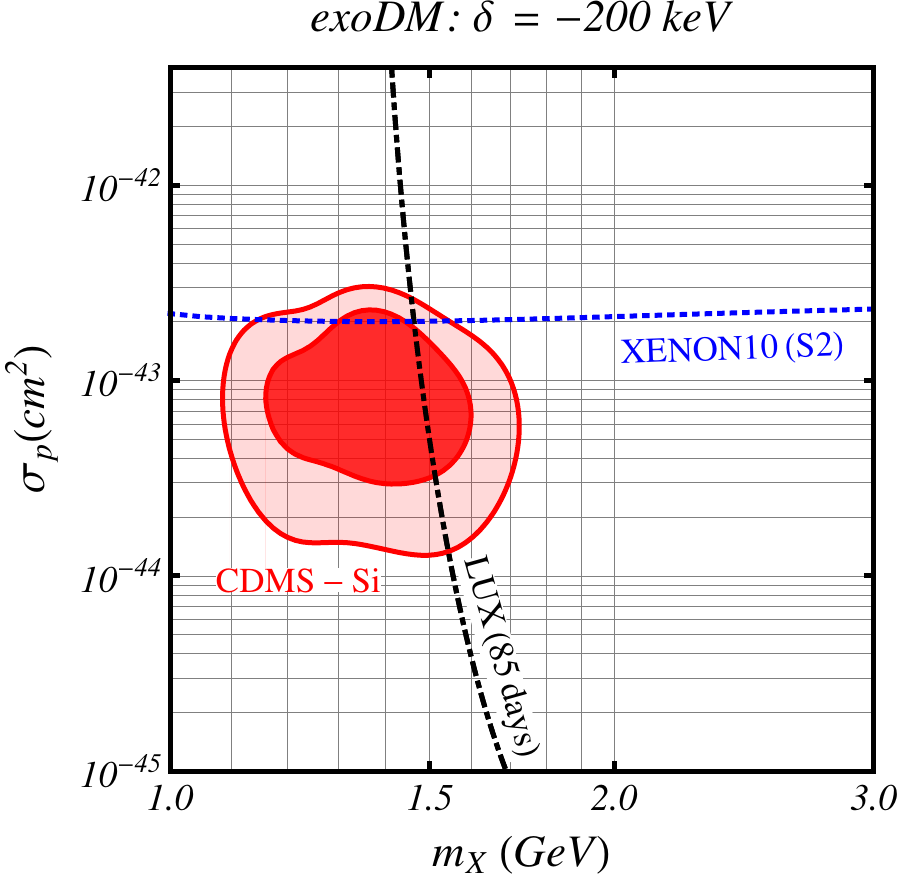}
                                  
   \caption{In each panel we fix the mass-splitting $\delta \equiv m_{2} - m_{1}$ between the two DM states, to $\delta =-50$ keV (left panel) and $\delta = -200$ keV (right panel), whilst treating the DM-proton cross section, $\sigma_{p}$, and DM mass, $m_{X_1}$, as free parameters to be fit by the data. The 90$\%$ CL limits from LUX (black dot-dashed) and XENON10 (blue dotted) are depicted along with the $68\%$ (dark red) and $90\%$ CL (light red) CDMS-Si best-fit regions. }
  \label{fig:compexoDM}
\end{figure*}

In Fig.~\ref{fig:compexoDM} we show two examples of exothermic DM parameter points ~\cite{ArkaniHamed:2008qn,Batell:2009vb,Graham:2010ca,Essig:2010ye,McCullough:2013jma}. In particular the right panel with $\delta = -200$ keV improves the compatibility between CDMS-Si and the null results of other searches with respect to ordinary ($\delta =0$) fits. 
%
We have found that LUX~\cite{Akerib:2013tjd} and the S2-only search by XENON10~\cite{Angle:2011th} are the most relevant searches constraining the parameter space. Ruling out this scenario for CDMS-Si requires a low threshold analysis: For example an updated S2-only analysis with high exposure from the XENON100 or LUX collaborations or by the proposed DAMIC-100 experiment, a very low threshold experiment with Silicon target material \cite{Aguilar-Arevalo:2013uua}.

%

\section{A Model of Excited State Dark Matter}
\label{model}
A number of models have been proposed for inelastic dark matter~\cite{TuckerSmith:2001hy} in the literature~~\cite{TuckerSmith:2002af,TuckerSmith:2004jv,Cui:2009xq,Baumgart:2009tn,Batell:2009vb,McCullough:2013jma}. The mass splitting $\delta$ can arise at tree-level from new Yukawa interactions splitting the components of a Dirac fermion into Majorana mass eigenstates, e.g \cite{Batell:2009vb}, or splitting otherwise mass degenerate Dirac fermions \cite{McCullough:2013jma}.
The mass splitting can also arise radiatively due to new non-abelian gauge bosons as in e.g. \cite{Baumgart:2009tn}.
We remain agnostic about the origin of the mass difference.  Two elements which the model must satisfy however are: (1) that sufficiently large cross section be generated to account for the results in Fig. 3; and (2) that the relic abundance of both $X_1$ and $X_2$ be accounted for.

We briefly review the basic properties of a model with Majorana DM and scattering that is dominantly inelastic with the required properties to describe the CDMS-Si data.
We take DM to be fermionic with Dirac mass $M$ and Majorana mass $ \delta \ll M$. In terms of the interaction eigenstate Dirac fermion $\chi$ we can write the mass terms as  
\be
\mathcal{L}_{\rm mass} = - M\bar{X}X- \frac{\delta}{4} \left(\bar{\chi^c}P_LX+ \bar{\chi^c}P_RX  \right) + \rm{h.c.}
\ee
In addition, we assume that the DM $\chi$ (interaction eigenstate) is coupled to a new gauge field $\phi_{\mu}$ of a spontaneously broken gauged $U(1)_{X}$. Now one can diagonalize the mass matrix and write the Lagrangian in terms of mass eigenstates $\chi_{1,2}$, 
\be \mathcal{L}_{\rm int} = \overline{X} i \gamma^{\mu} \left( \partial_{\mu} + i g_{X} \phi_{\mu} X\right) \longrightarrow
\overline{X}_{1} i \gamma^{\mu}  \partial_{\mu}X_1  + \overline{X}_{\,2} i \gamma^{\mu}  \partial_{\mu}X_2 + (i g_{X} \phi_{\mu} \overline{X}_{\,1} \gamma^{\mu}X_2 + {\rm h.c.}) + \mathcal{O}\left(\frac{g_X \delta }{M}\right),
\ee
where the mass eigenstates, $M_{1,2} = M \pm \delta$. From the above expression, it is thus clear that the interaction is dominantly off-diagonal.~\footnote{The next term in the $m/M$ expansion is diagonal but spin-dependent.}

The massive vector field $\phi_{\mu}$ can mix kinetically with the photon and the $Z$ boson via $\varepsilon F_X^{\mu \nu} B^{\mu \nu}$, where $F_X^{\mu \nu}$ is the field strength tensor of the $U(1)_{X}$ gauge boson, giving rise to a nonzero scattering cross section on nuclei. For direct detection phenomenology the interaction through photon mixing is dominant. 


 For $m_{X_i} \gtrsim m_{\phi}$ the annihilation is dominated by the channel into $\phi_{\mu}$ pairs, with the cross section
 \be \langle \sigma_{\bar{X}_iX_i \rightarrow \phi \phi} v \rangle = \frac{\pi \alpha_{X}^{2}}{m_{X_i}^{2}} \sqrt{1 - \left(\frac{m_{\phi}^{2}}{m_{X_i}^{2}}\right)}\ , \quad i=1,2\ ,  
 \ee
 where $\alpha_{X} \equiv g_{X}^{2}/4\pi$ and $g_{X}$ is the DM-dark photon gauge coupling. We see for example that for $m_{X_1} = 1$ GeV we need $\alpha_{X} = 7\times 10^{-5}$. 

Simultaneously we must also explain the observed scattering on nuclei. This proceeds through the kinetic mixing term with a cross section
\be \sigma_{p} \simeq 16 \pi \alpha_{X} \alpha_{EM} \, \varepsilon^{2}  \,\frac{\mu_{p}^{2}}{m_{\phi}^{4}}
\ee
where $\mu_{p}$ is the DM-proton reduced mass, and $\alpha_{EM}$ is the electromagnetic fine-structure constant. We have kept only the (by far) dominant contribution from the photon exchange.

\begin{figure*}[t!]
  \centering
 \includegraphics[width=0.40\textwidth]{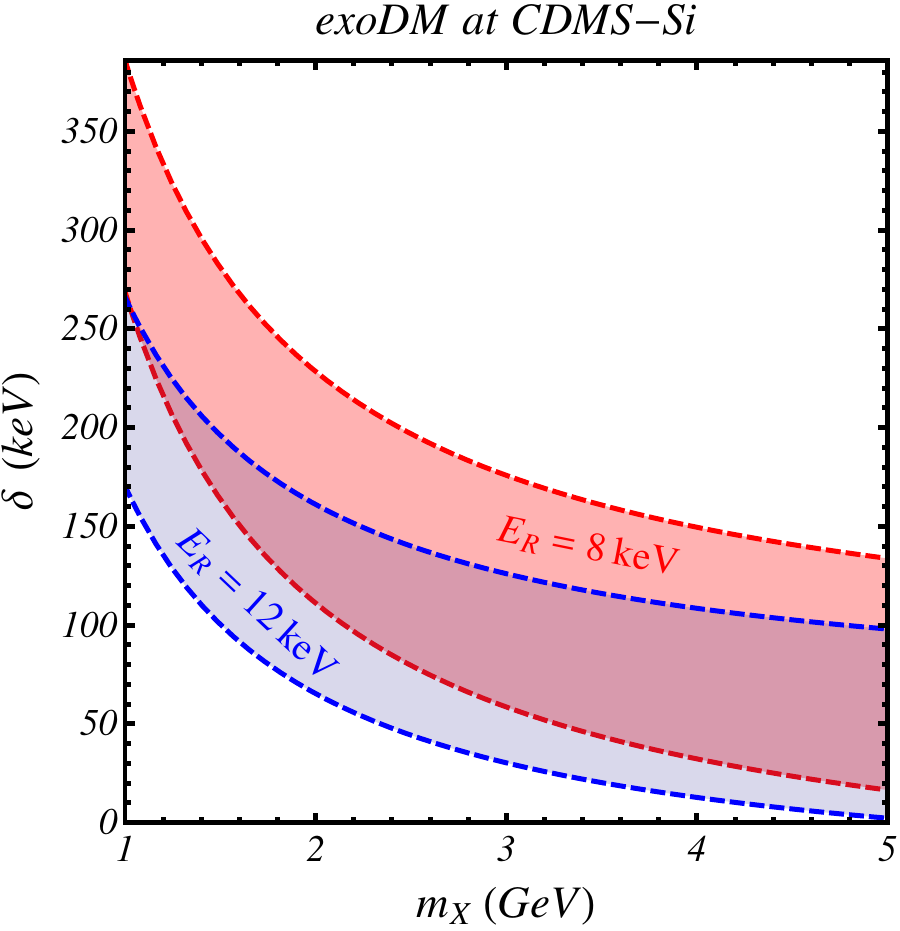}
  \includegraphics[width=0.45\textwidth]{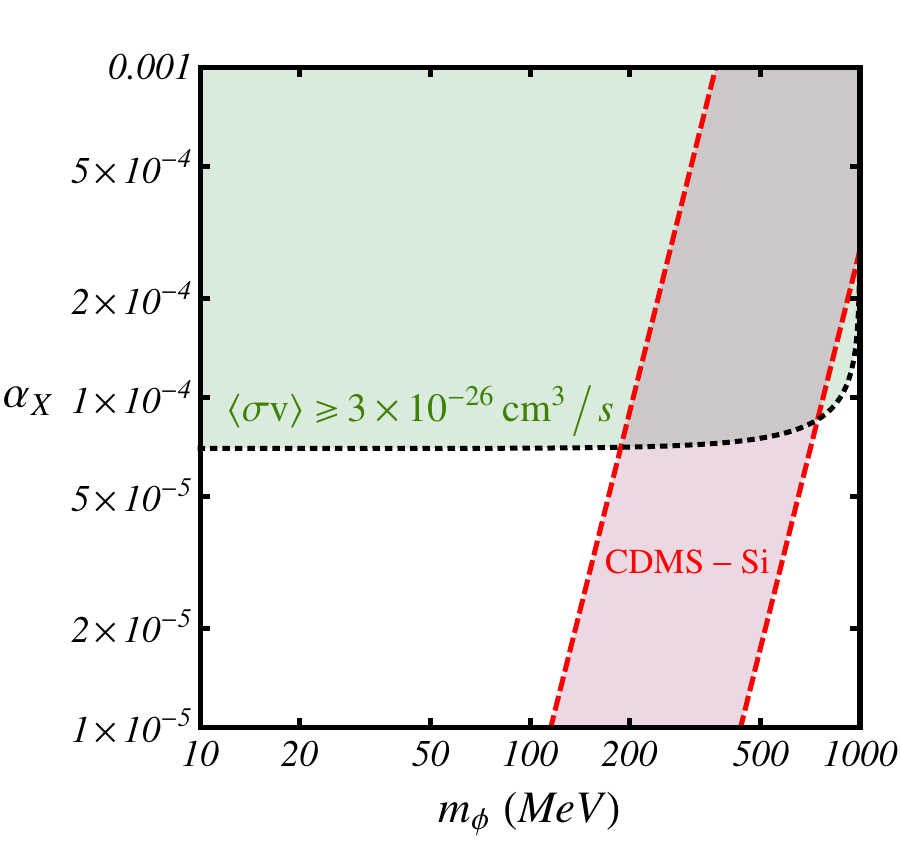}
   \caption{{\it Left:} Values of the mass splitting $\delta$ and DM mass $m_{X}$ that can account for the lowest and highest energy recoil events at CDMS-Si. {\it Right:} Region in the $(\alpha_{X},m_{\phi})$ plane accommodating the CDMS-Si and relic abundance constraints with fixed kinetic mixing, $\varepsilon = 10^{-6}$ and DM mass, $m_{X} = 1$ GeV. The correct relic abundance is obtained for (a)symmetric DM for parameter points on the dashed green curve (in the shaded green region). }
  \label{fig:relic}
\end{figure*}

Implicit in the above, is the assumption that a significant population of the excited state $X_1$ remains at present times. Indeed, the excited state can both decay into the ground state, $X_1 \rightarrow $ SM $+X_2$, or self-annihilate into two ground states $X_1 X_1\rightarrow X_2X_2$. We will largely follow the estimates in~\cite{Batell:2009vb}.

The decay of the excited state proceeds dominantly into $X_1 \rightarrow X_2 + 3 \gamma$, via an $e^{+}e^{-}$ loop. This decay rate has been estimated to be~\cite{Pospelov:2008jk,Batell:2009vb}
\be \Gamma_{X_1 \rightarrow X_2 + 3 \gamma} = 2\times 10^{-50}\,{\rm GeV} \times
\left(\frac{\varepsilon}{10^{-6}}\right)^2 
\left(\frac{\alpha_{X}}{10^{-3}}\right)
\left(\frac{\delta}{200\,{\rm keV}}\right)^{13}\left(\frac{200\,{\rm MeV}}{m_\phi}\right)^4, 
\ee
where we have chosen $\varepsilon = 10^{-6}$ such that the relic abundance and CDMS-Si favored regions overlap in Fig.~\ref{fig:relic}. Thus we see that with these parameters, the lifetime of $X_1$ is $\sim 10^{9}$ Gyr. 

The other potential worry is that the $X_1X_1 \rightarrow X_2 X_2$ scattering mediated by the $t$-channel exchange of a $\phi^{\mu}$ would be efficient enough to convert the excited states into ground states. This can happen if this conversion process stays in equilibrium below temperatures of order the mass-splitting $\delta$. To see that this does not occur, let us briefly sketch the thermal history of the model. After chemical decoupling, the DM density becomes fixed but it retains kinetic equilibrium with the SM by elastically scattering on electrons. Estimating the rate of this scattering as 
\be
\Gamma_{X_{i} e} \simeq \varepsilon^{2} \frac{\alpha_{X}\alpha_{EM}}{m_{\phi}^{4}} T^{5}
\ee
we find that this fails to keep up with Hubble expansion and hence decouples around $T^{*} \sim 20$ MeV. After this point, the DM sector, consisting of $X_1$ and $X_2$, has a temperature $T_{X} = T_{\gamma}^{2} /T^{*}$ where $T_{\gamma}$ is the SM temperature. Now we can estimate the temperature at which $X_1X_1 \rightarrow X_2 X_2$ freezes out by equating $H(T_{\gamma}) = n_{X_1}(T_{X}) \langle \sigma_{X_1X_1 \rightarrow X_2 X_2} v \rangle$. With $ \langle \sigma_{X_1X_1 \rightarrow X_2 X_2} v \rangle \simeq \pi \frac{\alpha_{X}^{2}m_{X}^{2}} {m_{\phi}^{4}}$ we find that the conversion process falls out of equilibrium also around 20 MeV. Thus we expect no large depletion of excited states and roughly estimate $n_{X_1} \approx n_{X_2}$. Therefore for consistency one should rescale the cross sections in Fig.~\ref{fig:compexoDM} by a factor of 2 since the density of $X_1$ is half that of the total DM density. 

\section{Conclusions}
We have discussed dark matter that inelastically scatters and de-excites in direct detection experiments (``Exothermic'' DM) as an explanation for the CDMS-Si events, consistent with all null searches. The constraints from LUX and XENON10 require the mass-splitting to be $|\delta| \gtrsim 50$ keV as already discussed in \cite{Frandsen:2013cna}. At the same time, the CDMS-Si data itself only allow splittings up to  $|\delta|  \sim 200$ keV since for larger splittings one cannot simultaneously account for the observed low and high energy events. 

In a simple model with a kinetically mixed dark photon, we have shown that the CDMS-Si rate and the thermal relic abundance can be accommodated. To rule out this scenario with a Xenon-based experiment will require an S2-only analysis (or other low-threshold technique~\cite{Arisaka:2012ce}) with high exposure.

\section*{Note Added}
While this manuscript was being prepared for upload the paper~\cite{Fox:2013pia} appeared which also considers LUX constraints on exothermic DM as an explanation of CDMS-Si data. The authors perform a global fit to the 3 parameters $\sigma_p, m_X$ and $\delta$. We find similar limits in parameter space, see e.g. Fig.~\ref{fig:relic}. 
\acknowledgements
The CP3
-Origins centre is partially funded by the Danish National Research Foundation, grant
number DNRF90. MTF acknowledges a ÔSapere AudeÕ Grant no. 11-120829 from the
Danish Council for Independent Research. 

\bibliographystyle{ArXiv}
\bibliography{sexy.bib}

\end{document}